\documentclass[conference, letterpaper]{IEEEtran2014}

\usepackage[latin1]{inputenc}
\usepackage{graphicx}
\usepackage{amsthm}
\usepackage{amssymb}
\usepackage{amsmath}
\usepackage{cite}
\usepackage{url}
\usepackage{balance}
\usepackage{array}
\usepackage{booktabs}

\usepackage{algorithm}
\usepackage{algpseudocode}

\usepackage{multirow}

\usepackage{fancyhdr}
\usepackage{datetime}

\newcounter{MYtempeqncnt}

\usepackage{mathtools}

\usepackage{epstopdf}

\newtheorem{myFact}{Fact}

\allowdisplaybreaks  

\IEEEoverridecommandlockouts

\def \Nf                {N_f} 

\begin{document}

\title{Beamforming Codebook Compensation for Beam Squint with Channel Capacity Constraint}

\author{\IEEEauthorblockN{Mingming Cai, J. Nicholas Laneman and Bertrand Hochwald\\}
\IEEEauthorblockA{Department of Electrical Engineering, University of Notre Dame\\
Email: \texttt{\{mcai, jnl, bhochwald\}@nd.edu}}
}

\maketitle

\begin{abstract}
Analog beamforming with phased arrays is a promising technique for 5G wireless communication in millimeter wave bands. A beam focuses on a small range of angles of arrival or departure and corresponds to a set of fixed phase shifts across frequency due to practical hardware constraints. In switched beamforming, a discrete codebook consisting of multiple beams is used to cover a larger angle range. However, for sufficiently large bandwidth, the gain provided by the phased array is frequency dependent even if the radiation pattern of the antenna elements is frequency independent, an effect called beam squint. This paper shows that the beam squint reduces channel capacity of a uniform linear array (ULA). The beamforming codebook is designed to compensate for the beam squint by imposing a channel capacity constraint. For example, our codebook design algorithm can improve the channel capacity by 17.8\% for a ULA with 64 antennas operating at bandwidth of 2.5~GHz and carrier frequency of 73~GHz. Analysis and numerical examples suggest that a denser codebook is required to compensate for the beam squint compared to the case without beam squint. Furthermore, the effect of beam squint is shown to increase as bandwidth increases, and the beam squint limits the bandwidth given the number of antennas in the array.
\end{abstract}

\section{Introduction}
\label{sec:introduction}
Millimeter-wave (mmWave) bands can provide the opportunity for GHz of spectrum in next-generation (5G) cellular communication \cite{roh2014millimeter}.  However, the signals at mmWave frequencies experience significantly higher path loss, typically 20~dB or more, than signals below 6~GHz \cite{rappaporttheodores2002}. Beamforming with a phased array using a large number of antennas can compensate for the attenuation of mmWave while supporting mobile applications \cite{heath2015overview}. In this paper, we consider switched analog beamforming with one phased array and associated RF chain for mobile communication. To cover a certain range of angle of arrival (AoA) or angle of departure (AoD) in switched beamforming, a ``codebook'' consisting of multiple beams is required \cite{song2015codebook}, with each codeword in the codebook consisting of a set of beamforming phases.



Typically, each branch in a phased array has the same phase shift for all frequencies within the band of study, and it is a relatively good approximation to a time delay for narrowband signals. However, this approximation breaks down for signals with large bandwidth if the angle of arrival (AoA) or angle of departure (AoD) is away from broadside, because phase shifters correspond to different time delays for different frequencies in the band. The net result is that the array response varies over frequency, and the beams for frequencies other than the carrier ``squint'' as a function of frequency \cite{mailloux2005phased}. This phenomenon is called beam squint \cite{SeyedGarakkoui2011BeamSquinting}. Beam squint introduces array gain variations over frequency for a given AoA or AoD \cite{cai2016effect}. Approaches to eliminate or reduce beam squint have been studied, including true time delay \cite{longbrake2012true}. However, such approaches are not appealing for mobile wireless communication due to combinations of high implementation cost, significant insertion loss, large size, or excessive power consumption \cite{cai2016effect, longbrake2012true}.

In this paper, we study three aspects of beam squint. First, we show several effects of beam squint on channel capacity and spectral efficiency. For example, beam squint reduces the channel capacity, especially if the number of antennas and fractional bandwidth, defined as the ratio of the bandwidth to the carrier frequency, are large. Second, a beamforming codebook is designed to compensate for the beam squint by imposing a channel capacity constraint. Third, it is illustrated that the required codebook size increases with the number of antennas in the array as well as the fractional bandwidth. There is also a tight upper bound for the fractional bandwidth with fixed number of antennas; beyond this upper bound on the fractional bandwidth, no codebook satisfying the constraint exists.

\section{System Model}
\begin{figure}
\begin{center}
\includegraphics[width=72 mm]{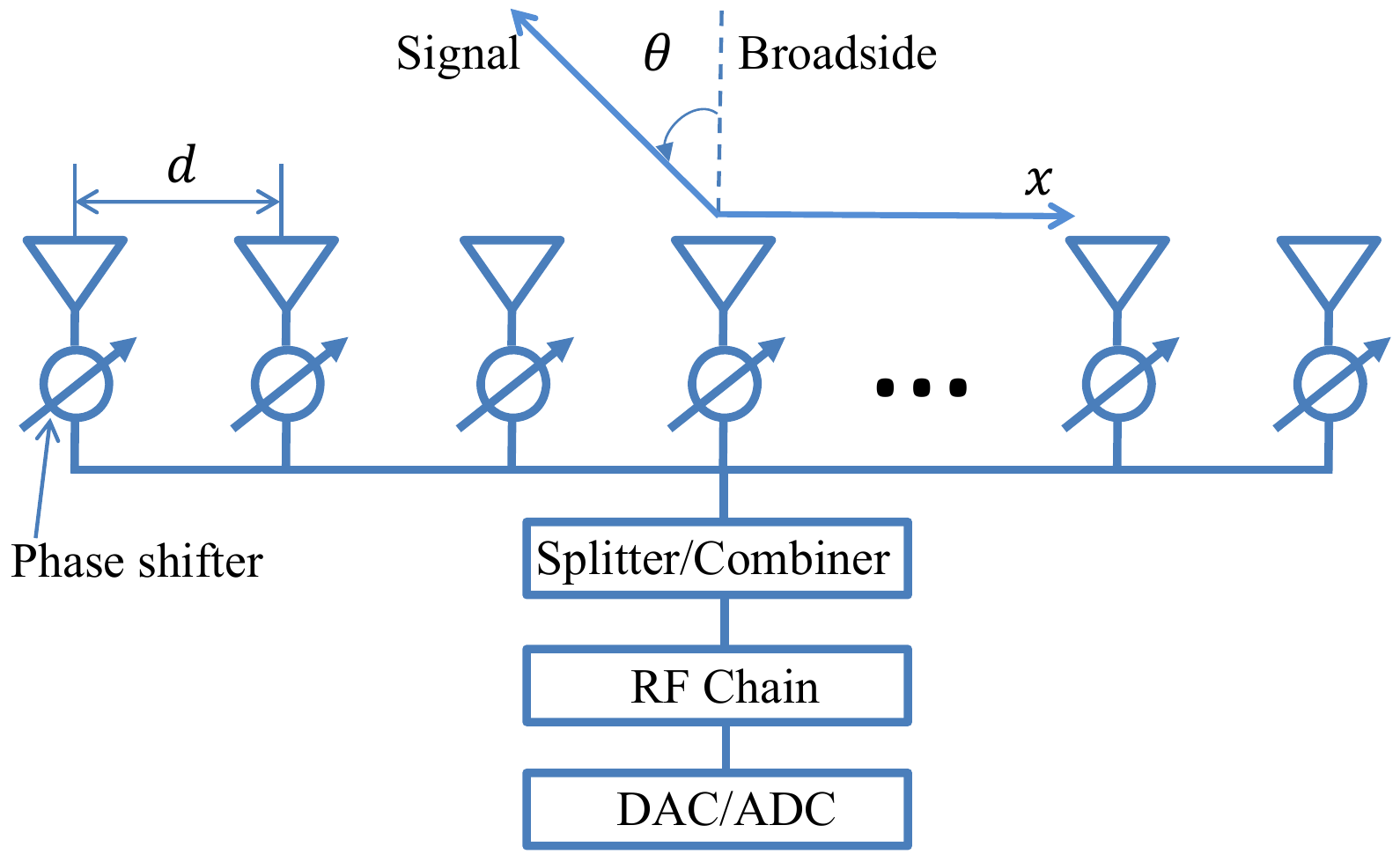}
\end{center}
\caption{Structure of a uniform linear array (ULA) with analog beamforming using phase shifters.  The distance between the adjacent antennas is $d$, and $\theta$ denotes either the angle-of-arrival (AoA) for reception or the angle-of-departure (AoD) for transmission.}
\label{fig:array-structure}
\end{figure}
We consider an OFDM system with bandwidth $B$ and $\Nf$ subcarriers. Label the subcarriers as $0,1,\dots,N_f-1$ with increasing frequency. A uniform linear array (ULA) with $N$ identical and isotropic antennas is used as shown in Fig.~\ref{fig:array-structure}. The distances between two adjacent antenna elements are all $d$. The AoD or AoA $\theta$ is the angle of the signal relative to the array broadside, increasing counterclockwise. Similar to \cite{cai2016effect}, to better study the beam squint, angle is denoted in a virtual angle domain, defined as
\begin{align}
\psi=\sin \theta.
\end{align}

Let the beamforming vector be denoted as
\begin{align}
{\mathbf{w}} = \left[ {e^{j\beta _1},\dots,e^{j\beta_n},\dots,e^{j\beta_N}} \right]^T,
\label{eq:beam-forming-vector}
\end{align}
where $\beta_n$ is the phase of $n$th phase shifter connected to the $n$th antenna element. For simplicity, we assume phase shifters have infinite resolution, i.e., $\beta_n$, $n=1,2,...,N$ is continuous. To maximize the array gain in a direction $\psi_F$, signals from or to all $N$ antennas should have the same time delay by setting up phased shifters. From \cite{cai2016effect}, to focus the array for the carrier frequency $f_c$ on direction $\psi_F$, we have
\begin{align}
{\beta _n} \left( \psi_F \right)=  2\pi {\lambda_c^{ - 1}} d \left( {n - 1} \right) \psi_F, \ n=1,2,\dots,N,
\label{eq:phases-required-for-fine-beams}
\end{align}
where $\lambda_c$ is the wavelength of the carrier frequency. We call the beam with phase shifts described in (\ref{eq:phases-required-for-fine-beams}) a \emph{fine beam}, in which the array focuses on a single AoA/AoD, maximizing the array gain. We study fine beams throughout the paper.



\subsection{Model for Beam Squint}
The phase shifts are fixed for all frequencies within bandwidth $B$, introducing beam squint. Denote a subcarrier's absolute frequency $f$ relative to the carrier frequency $f_c$ as
\begin{align}
f=\xi f_c,
\label{eq:freq-coefficient}
\end{align}
where $\xi$ is the ratio of the subcarrier frequency to the carrier frequency. Thus, $f \in \left[ f_c-\frac{B}{2}, f_c+\frac{B}{2} \right]$. Define the fractional bandwidth
\begin{align}
b:={B}/{f_c}.
\label{eq:fractional-bandwidth}
\end{align}
Then, $\xi \in \left[1-b/2, 1+b/2\right]$. For instance, $\xi$ varies from 0.983 to 1.017 and $b=0.034$ for a system with 2.5~GHz bandwidth at 73~GHz carrier frequency.

\begin{figure}
\begin{center}
\includegraphics[width=88 mm]{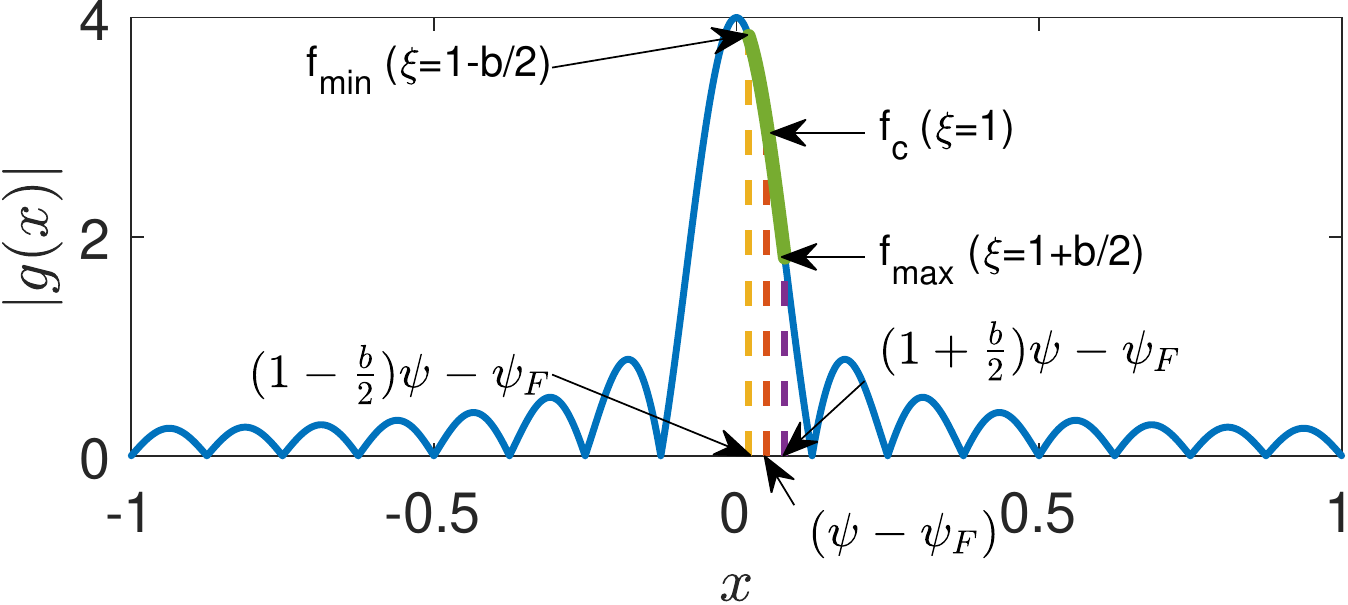}
\end{center}
\caption{Example of $g \left(x \right)$ from (\ref{eq:ms-array-gain-one-parameter}) for a fine beam with $N=16$ antennas and half-wavelength antenna spacing $d=\lambda_c/2$. The curve in bold illustrates the array gain variations for different frequencies within the band.}
\label{fig:Beam_shape_plot_fine_beam_psi}
\end{figure}

Assume $d=\lambda_c/2$. From \cite{cai2016effect}, the array gain for subcarrier $\xi$, AoA/AoD $\psi$, and beam focus angle $\psi_F$ can be expressed as $g \left(\xi \psi - \psi_F\right)$, where
\begin{align}
g \left(x \right) = \frac{{\sin \left( {\frac{{N\pi x}}{2}} \right)}}{{\sqrt N \sin \left( {\frac{\pi x }{2}} \right)}} e^{j \frac{\left(N-1\right) \pi x} {2}}.
\label{eq:ms-array-gain-one-parameter}
\end{align}
The gain for subcarrier $\xi$ at AoA/AoD $\psi$ is equivalent to the gain for the carrier frequency at AoA/AoD $\psi^{\prime}=\xi \psi$. The effect of beam squint increases if the AoA/AoD is away from the beam focus angle. Fig.~\ref{fig:Beam_shape_plot_fine_beam_psi} illustrates an example of array gains for different frequencies as well as the plot of $\left|g \left(x \right) \right|$. The main lobe is the one with the highest gain, and the side lobes have smaller gains. Our study focuses on the main lobe. The range of the main lobe of $g \left(x \right) $ is $x \in \left[-\frac{2}{N}, \frac{2}{N} \right]$, and the span of the main lobe is $\frac{4}{N}$. If $x \in \left[-\frac{4}{N \pi}, \frac{4}{N \pi}\right]$, $g\left(x\right)$ is strictly concave. The maximum array gain is
\begin{align}
g_m=\mathop {\max }\limits_{\psi \in \left[-1,1\right]}  g \left({\psi-\psi_F}\right) =\sqrt{N}.
\label{eq:maximum-array-gain}
\end{align}

\subsection{Channel Capacity of OFDM system}
For simplicity, we only study the beam squint of a single array. To be specific, assume there is no beam squint at the transmitter and equal power is allocated to all subcarriers in the transmitter, which is common if channel state information is not known at the transmitter. In mmWave bands with fine beams at both the transmitter and receiver, the channel becomes sparse \cite{heath2015overview}. We therefore assume there is only one signal path, so that the channel is flat over frequency.

The total power received within bandwidth $B$ by one antenna before beamforming is $P$, and the spectral density of white Gaussian noise modeling thermal noise is $\sigma^2$. The channel capacity with beam squint at AoA $\psi$ and beam focus angle $\psi_F$ is
\begin{align}
C_{\text{BS}}\left( {{\psi_F},\psi, b} \right)
=\frac{B}{N_f} \sum\limits_{n = 0}^{{N_f} - 1} {\log \left( {1 + \frac{{P}{{\left| {g \left( { \xi_n \psi- \psi _F } \right)} \right|}^2}}{ B\sigma ^2}} \right) },
\label{eq:capacity-with-beam-squint-no-pilot}
\end{align}
where
\begin{align}
{\xi _n} = 1+\frac{\left( 2n-N_f+1 \right) b}{2N_f}.
\label{eq:xi-n}
\end{align}
For comparison, the channel capacity without beam squint at AoA $\psi$ and beam focus angle $\psi_F$ is
\begin{align}
C_{\text{NBS}}\left( {{\psi_F},\psi} \right) = {B\log \left( {1 + \frac{P {{\left| {g \left( {\psi-\psi _F} \right)} \right|}^2} }{B \sigma ^2}} \right)}.
\label{eq:capacity-no-beam-squint-no-pilot}
\end{align}
Again, the channel capacity without beam squint can be realized by replacing phase shifters with true-time-delay devices \cite{longbrake2012true}. However, the cost is too high for mobile communication. We define $C_{\text{NBS}}$ for comparison with $C_{\text{BS}}$. Both $C_{\text{NBS}}\left( {{\psi_F},\psi} \right)$ and $C_{\text{BS}}\left( {{\psi_F},\psi, b} \right)$ are maximized if $\psi=\psi_F$.

\section{Effect of Beam Squint on Channel Capacity}

The angle range with beam focus angle $\psi_F$ and array gain no less than $rg_m,$ $0<r<1$, can be expressed as
\begin{align}
 \mathcal{R}_0 \left( {\psi_F}, r \right) = \left\{ {\psi :\left| {g \left( {\psi-\psi _F } \right)} \right| \ge rg_m,\psi  \in \left[ -1, 1\right]} \right\}.
\end{align}
Define $\mathcal{R}_{3_{\text{dB}}} \left( {\psi_F}\right)=\mathcal{R}_0 \left( {\psi_F}, \frac{\sqrt 2}{2} \right) $ as the angle set for 3~dB beamwidth about beam focus angle $\psi_F$.

\begin{myFact}
If all the subcarriers lie within a range defined by $\left(1+\frac{b}{2}\right) \psi \le \left(\psi_F +\frac{4}{N \pi} \right)$ and  $\left(1-\frac{b}{2}\right) \psi \ge \left(\psi_F -\frac{4}{N \pi}\right)$, i.e., $\frac{\psi_F -\frac{4}{N \pi} }{\left(1-\frac{b}{2}\right)} \le \psi \le \frac{\psi_F +\frac{4}{N \pi} }{\left(1+\frac{b}{2}\right)} $, then
\begin{align}
C_{\text{BS}}\left( {{\psi_F},\psi , b } \right) \le C_{\text{NBS}}\left( {{\psi_F},\psi} \right).
\end{align}
Equality is achieved if and only if $\psi=0$.
\label{channel-capacity-degradation}
\end{myFact}
Fact~\ref{channel-capacity-degradation} can be proved with Jensen's Inequality and the concavity of $g\left(x\right)$ for $x\in  \left[-\frac{4}{N\pi}, \frac{4}{N\pi} \right]$. The proof is omitted.


The angle range defined by $\frac{\psi_F -\frac{4}{N \pi} }{\left(1-\frac{b}{2}\right)} \le \psi \le \frac{\psi_F +\frac{4}{N \pi} }{\left(1+\frac{b}{2}\right)} $ is much larger than that of the 3 dB beamwidth defined by $\mathcal{R}_{3_{\text{dB}}}\left( {{\psi _F}} \right)$.

The spectral efficiency with beam squint can be defined as
\begin{align}
E _{\text{BS}}\left( {{\psi_F},\psi, b} \right):&=\frac{C_{\text{BS}}\left( {{\psi_F},\psi, b} \right)}{B}.
\end{align}
We also have the following Fact.
\begin{myFact}
Fix the ratio $\frac{P}{B \sigma^2}$. If fractional bandwidth $b_1<b_2$, and $\frac{\psi_F -\frac{4}{N \pi} }{\left(1-\frac{b_2}{2}\right)} \le \psi \le \frac{\psi_F +\frac{4}{N \pi} }{\left(1+\frac{b_2}{2}\right)}$,
\begin{align}
{E_{\text{BS}}\left( {{\psi_F},\psi , b_1 } \right)}  \ge  { E_{\text{BS}}\left( {{\psi_F},\psi ,b_2 } \right)}.
\end{align}
Equality holds if and only if $\psi=0$.
\label{spectrum-effeciency-decrease-as-fractional-bandwidth}
\end{myFact}
Fact~\ref{spectrum-effeciency-decrease-as-fractional-bandwidth} can also be proved with Jensen's Inequality and the concavity of $g\left(x\right)$ for $x\in  \left[-\frac{4}{N\pi}, \frac{4}{N\pi} \right]$. The proof is omitted.

By fixing the ratio $\frac{P}{B \sigma^2}$, the transmit power varies linearly with $B$ for fixed noise spectral density $\sigma^2$. As the fractional bandwidth increases, the spectral efficiency decreases according to Fact~\ref{spectrum-effeciency-decrease-as-fractional-bandwidth}.

\begin{figure}
\centering
\includegraphics[width=88 mm]{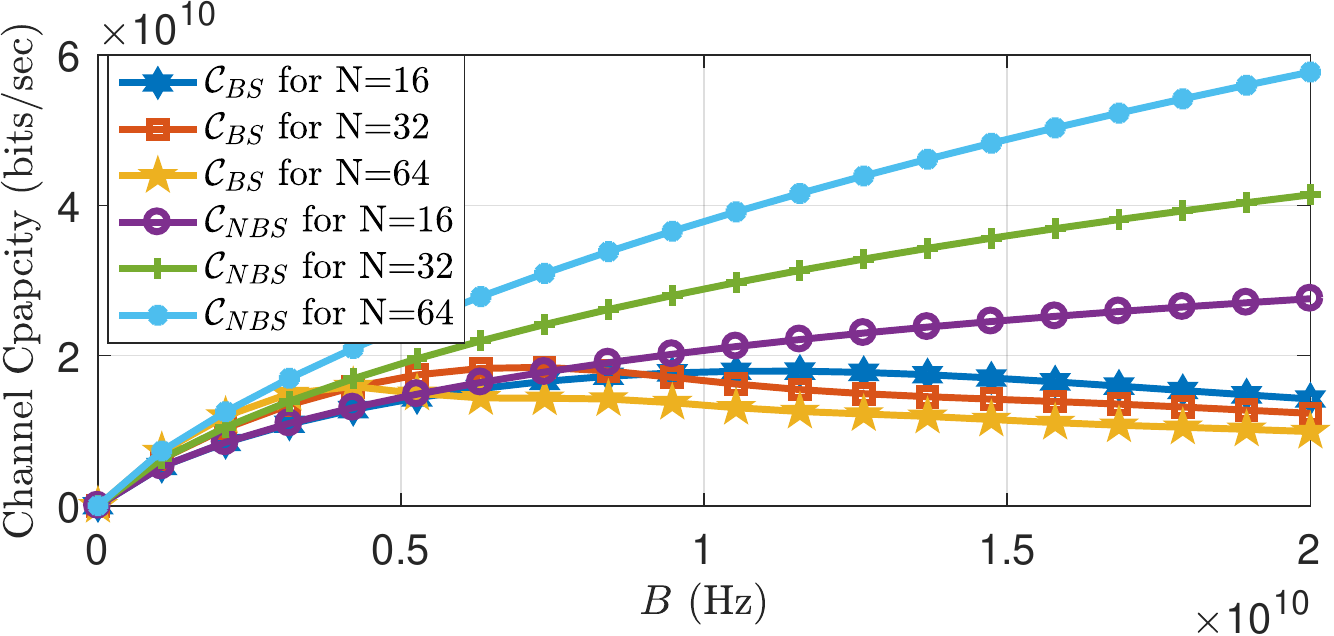}
\caption{Capacity with beam squint $C_{\text{BS}}\left( {{\psi_F},\psi, b} \right)$ and without beam squint $C_{\text{NBS}}\left( {{\psi_F},\psi, b} \right)$ as a function of bandwidth for $\psi=\psi_F=0.9$ with fixed ${P/\sigma^2}=2\times 10^9$ Hz, $d=\lambda_c/2$, and $N_f=2048$.}
\label{fig:LA2_Channel_capacity_vsB_multipleN}
\end{figure}

If the fractional bandwidth increases while $P$ remains the same by having the same transmit power, $C_{\text{NBS}}\left( {{\psi_F},\psi} \right)$ increases with $B$ \cite{cover2012elements}. Let $B \to \infty$,  then $C_{\text{NBS}}\left( {{\psi_F},\psi} \right) \to \frac{P {{\left| {g \left( {\psi-\psi _F} \right)} \right|}^2} }{\sigma ^2} \log_2{e}$ \cite{cover2012elements}. However, the increasing trend does not apply to $C_{\text{BS}}\left( {{\psi_F},\psi, b} \right)$. If $B$ is increased with fixed total transmit power, $C_{\text{BS}}\left( {{\psi_F},\psi, b} \right)$ has trends of first increasing and then decreasing. Fig.~\ref{fig:LA2_Channel_capacity_vsB_multipleN} illustrates three examples. For small $B$, the effect of beam squint is limited. $C_{\text{BS}}\left( {{\psi_F},\psi, b} \right)$ increases due to the dominant effect of the growing bandwidth, similar to $C_{\text{NBS}}\left( {{\psi_F},\psi} \right)$. As $B$ further increases, the effect of beam squint dominates, and a larger portion of the subcarriers have small array gains. Some subcarriers may even fall outside the main lobe. Therefore, $C_{\text{BS}}\left( {{\psi_F},\psi} \right)$ begins to decrease.


\section{Beamforming Codebook Design}
To compensate for beam squint, the beamforming codebook can be designed to satisfy a minimum channel capacity constraint for each beam. Suppose we want the codebook to cover the angle range $\left[ { - {\psi _m},{\psi _m}} \right]$.
The coverage of a fine beam $\mathbf{w}$ with beam focus angle $\psi_F(\mathbf{w})$ and target rate $C_t$ is the set
\begin{align}
\mathcal{R}_c \left( \mathbf{w}, C_t \right):=&\left\{ \psi \in \left[ -\psi_m, \psi_m \right]: C_{\text{BS}}\left( {{\psi_F\left(\mathbf{w}\right)},\psi, b} \right) \ge {C_t} \right\}.
\label{beam-range-definition-minimum-capacity}
\end{align}
Selection of the capacity threshold ${C_t}$ depends on the system design requirements. Generally, we can select the channel capacity without beam squint as a benchmark for $C_t$. For example, select
\begin{align}
C_{t}\left( r \right) = {B\log \left( {1 + \frac{r^2NP }{B \sigma ^2}} \right)} =C_{\text{NBS}}\left( \psi_F, \max \mathcal{R}_0 \left( \psi_F, r \right) \right),
\label{eq:typical-capacity-threshold}
\end{align}
where $0<r<1$. A reasonable choice is $C_{t,3_{\text{dB}}}=C_{t}\left( \frac{\sqrt 2}{2} \right)$, the capacity without beam squint at the the 3~dB beamwidth, i.e., the array gain is 3~dB below $g_m$.

The codebook is defined as
\begin{align}
{\mathcal C}_c := \left\{ {   {{\bf{w}}_1}, {{\bf{w}}_2}, \dots, {{\bf{w}}_M}: \bigcup\limits_{i = 1}^M {{\mathcal{R}_c}\left( {{\mathbf{w}}_i, C_t} \right)} \supseteq \left[ { - {\psi _m},{\psi _m}} \right]} \right\}.
\label{codebook-definition-minimum-capacity}
\end{align}
The $i$th beam in the codebook has weights $\mathbf{w}_i$ and beam focus angle $\psi_{i,F}$. We further assume the beam focus angles satisfy
\begin{align}
-\psi_m \le \psi_{1,F}< \psi_{2,F} < \dots< \psi_{M,F} \le \psi_m.
\end{align}

The codebook size $\left| \mathcal{C}_c \right|=M$. Let $\mathbf{\Omega}_c$ be the set of all codebooks that meet the requirements of (\ref{codebook-definition-minimum-capacity}) for fixed values of $N$, $b$, $C_t$, $\frac{P}{B\sigma^2}$, and $\psi_m$. There is an infinite number of codebooks in $\mathbf{\Omega}_c$. Among them, we focus on the ones that achieve the minimum codebook size. The problem that needs to be solved is to find
\begin{align}
\mathcal{C}_{min}=\mathop {\arg \min }\limits_{\mathcal{C}_c \in \mathbf{\Omega}_c}  \quad   \left| \mathcal{C}_c \right|. \label{codebook-design-minimum-channel-capacity-formulation}
\end{align}

\subsection{Analysis of the Codebook}
Define left and right edges of the coverage of beam $i$ as
\begin{align}
&\psi_{i,l}=\min {{{\mathcal{R}_c}\left( {{\mathbf{w}}_i, C_t} \right)}}, \\
&\psi_{i,r}=\max {{{\mathcal{R}_c}\left( {{\mathbf{w}}_i, C_t} \right)}},
\end{align}
respectively. The corresponding beamwidth is
\begin{align}
\Delta \psi_i=\psi_{i,r}-\psi_{i,l}.
\end{align}
To achieve the minimum codebook size, there should be as little overlap as possible in the coverages for all the beams. One solution is
\begin{align}
{{\mathcal{R}_c}\left( {{\mathbf{w}}_i, C_t} \right)} \cap {{\mathcal{R}_c}\left( {{\mathbf{w}}_j, C_t} \right)}=\varnothing, \quad \text{for }i \neq j.
\end{align}
Therefore,
\begin{align}
\psi_{i,r}=\psi_{i+1,l},\quad i=1,2,\dots, M-1.
\end{align}
The codebook can be designed by aligning the beams one by one. $\psi_{i,l}$ and $\psi_{i,r}$ can be derived from $\psi_{i,F}$ by solving
\begin{align}
C_{\text{BS}}\left( {{\psi_{i,F}}, \psi, b} \right)=C_t.
\end{align}
As there is not a closed-form solution for this equations, numerical methods must be applied. Take $\psi_{i,r}$ for example. First, we find the beamwidth for $C_t$ without beam squint, $\Delta\psi_{C_t}$, which is the difference between the two solutions of
$C_{\text{NBS}}\left( {{\psi_F},\psi} \right)=C_t$. All beams without beam squint have the same beamwidth. From Fact~\ref{channel-capacity-degradation}, $\psi_{i,r}$ must follow
\begin{align}
\psi_{i,F} \le \psi_{i,r}<\psi_{i,F}+\frac{\Delta\psi_{C_t}}{2}.
\end{align}
As $\psi$ increases from $\psi_{i,F}$ to $\psi_{i,F}+\frac{\Delta\psi_{C_t}}{2}$, $C_{\text{BS}}\left( {{\psi_{i,F}}, \psi, b} \right)$ monotonically decreases. Therefore, $\psi_{i,r}$ can be found through binary search in $\left[\psi_{i,F}, \psi_{i,F}+\frac{\Delta\psi_{C_t}}{2}\right)$ \cite{thomas2001introduction}. Similarly, $\psi_{i,l}$ can also be solved numerically. Suppose $\psi_{i,l}$ is known, $\psi_{i,F}$ must follow
\begin{align}
\psi_{i,l} \le \psi_{i,F}<\psi_{i,l}+\frac{\Delta\psi_{C_t}}{2},
\end{align}
where $C_{\text{BS}}\left( {{\psi_{i,F}}, \psi_{i,l}, b} \right)$ monotonically decreases as ${\psi_{i,F}}$ increases. Thus, $\psi_{i,F}$ can be solved numerically with binary search.

\subsection{Codebook Design}
\begin{algorithm}[t]
\caption{Codebook Design with Beam Squint}
\label{codebook-design-beam-squint}
\begin{algorithmic}[1]
\Procedure{1}{Odd-Number Codebook}
\State Let $k=1$
\State Align the first beam so that $\psi_{1,F}^{\prime}=0$
\State Derive $\psi_{1,r}^{\prime}$ from $\psi_{1,F}^{\prime}$
\While {$\psi_{k,r}^{\prime}<\psi_m$}
    \State Let $\psi_{k+1,l}^{\prime}=\psi_{k,r}^{\prime}$
    \State Derive $\psi_{k+1,F}^{\prime}$ from $\psi_{k+1,l}^{\prime}$
    \State Derive $\psi_{k+1,r}^{\prime}$ from $\psi_{k+1,F}^{\prime}$
    \State $\psi_{k+2,F}^{\prime}=-\psi_{k+1,F}^{\prime}$
    \State Align two beams with $\psi_{k+1,F}^{\prime}$ and  $\psi_{k+2,F}^{\prime}$
    \State $k=k+2$
\EndWhile
\State $\left| \mathcal{C}_{o} \right|=k$
\EndProcedure

\Procedure{2}{Even-Number Codebook}
\State Let $k=0$, $\psi_{0,r}^{\prime}=0$
\While {$\psi _{c,r}<\psi_m$}
    \State Let $\psi_{k+1,l}^{\prime}=\psi_{k,r}^{\prime}$
    \State Derive $\psi_{k+1,F}^{\prime}$ from $\psi_{k+1,l}^{\prime}$
    \State Derive $\psi_{k+1,r}^{\prime}$ from $\psi_{k+1,F}^{\prime}$
    \State $\psi_{k+2,F}^{\prime}=-\psi_{k+1,F}^{\prime}$
    \State Align two beams with $\psi_{k+1,F}^{\prime}$ and  $\psi_{k+2,F}^{\prime}$
    \State $k=k+2$
\EndWhile
\State $\left| \mathcal{C}_{e} \right|=k$
\EndProcedure
\State ${\left| \mathcal{C} \right| }_{min}=\min\left(\left| \mathcal{C}_{o} \right| , \left| \mathcal{C}_{e} \right| \right)$
\State Select the Procedure that achieves ${\left| \mathcal{C} \right| }_{min}$
\end{algorithmic}
\end{algorithm}

Among possible codebooks, we are more interested in the ones that are symmetric relative to the broadside. The beamwidth varies over beam focus angles because of beam squint. The codebook size is not known before beam alignment in the codebook. A codebook design algorithm is illustrated as shown in Algorithm~\ref{codebook-design-beam-squint}. Similar to the codebook algorithm in \cite{cai2016effect}, the codebook design is divided into two situations, odd codebook size ${\left| \mathcal{C} \right| }_{o}$ and even codebook size ${\left| \mathcal{C} \right| }_{e}$. The minimum codebook size ${\left| \mathcal{C} \right| }_{min}=\min\left(\left| \mathcal{C}_{o} \right| , \left| \mathcal{C}_{e} \right| \right)$. We need to note that no codebook exists if there is no solution in solving for $\psi_{i,r}$ or $\psi_{i,F}$.

\subsection{Constraint of $b$}
Fix the ratio $\frac{P}{B \sigma^2}$. As the fractional bandwidth $b$ increases, $\mathcal{R}_c \left( \mathbf{w}, C_t \right)$ decreases and the minimum codebook size increases. As $b$ increases further, $\mathcal{R}_c \left( \mathbf{w}, C_t \right)$ could be an empty set, that is, no solution exists in solving for $\psi_{i,r}$ given $\psi_{i,F}$ or solving for $\psi_{i,F}$ given $\psi_{i,l}$. Therefore, $b$ has a upper bound, denoted $b_{sup}$. As $b$ approaches $b_{sup}$, $\Delta \psi_1\to 0,$ $\Delta \psi_M\to 0$, and thus, ${\left| \mathcal{C} \right| }_{min} \to \infty$. The upper bound cannot be achieved, i.e., $b<b_{sup}$. Suppose $b=b_{sup}$ and $M$ is a finite number. Then, $\mathcal{R}_c \left( \mathbf{w}_M, C_t \right)=\psi_m$, and there will be a gap between $\mathcal{R}_c \left( \mathbf{w}_{M-1}, C_t \right)$ and $\mathcal{R}_c \left( \mathbf{w}_M, C_t \right)$. No codebook exists in this case.
%

$b_{sup}$ is also related to the number of antennas in the array.
\begin{myFact}
Design a codebook with Algorithm~\ref{codebook-design-beam-squint} and capacity threshold $C_{t}\left( r \right)$, $0<r<1$. With fixed $r$,
\begin{align}
b_{sup}\approx \frac{a}{N},
\end{align}
where $a$ is a constant.
\end{myFact}
For example, $b_{sup}\approx 3.04/N$ for $\frac{P}{B\sigma^2}=0$ dB,  $r=\frac{\sqrt2}{2}$ and $C_t=C_{t,3_{\text{dB}}}$.

\section{Numerical Results}
\label{sec:numerical-result}

\subsection{Improvement of Channel Capacity }
Algorithm~\ref{codebook-design-beam-squint} can provide a codebook with minimum channel capacity $C_{t} \left( r \right)$. A traditional codebook design algorithm allocates a beam to cover
$\psi \in \mathcal{R}_0 \left( {\psi_F}, r \right)$ and does not account for beam squint.  For $\psi$ in this set, the channel capacity would be ${C_{{\text{BS}}}}\left( \psi_F,\psi, b \right)$, and therefore minimum channel capacity
\begin{align}
{C_{\text{BS}, min}}\left( \psi_F, b, r\right)=\min\limits_{\psi \in  \mathcal{R}_0 \left( {\psi_F}, r \right)} {{C_{{\text{BS}}}}\left( \psi_F,\psi, b \right)}.
\end{align}
${C_{\text{BS}, min}}\left( \psi_F, b, r\right)$ is achieved if $\psi$ is the on the beam edge with larger magnitude.

Define the capacity improvement ratio as
\begin{align}
I\left( {{\psi_F},b, r} \right) = \frac{ C_t\left(r \right) - {C_{\text{BS}, min}}\left( \psi_F, b, r\right)} {{C_{\text{BS}, min}}\left( \psi_F, b, r\right)}.
\end{align}
For certain fractional bandwidth $b$, the maximum capacity improvement ratio for $\psi_F \in \left[-1,1\right]$ is
\begin{align}
I_{max}\left( b,r \right) = \mathop {\max }\limits_{{\psi_F} \in \left[ { - 1,1} \right]} {I\left( {{\psi_F},b,r} \right)}.
\end{align}
$I_{max}\left( b,r \right)$ is achieved if $\psi_F=\pm 1$.

\begin{figure}[t]
\centering
\includegraphics[width=88mm]{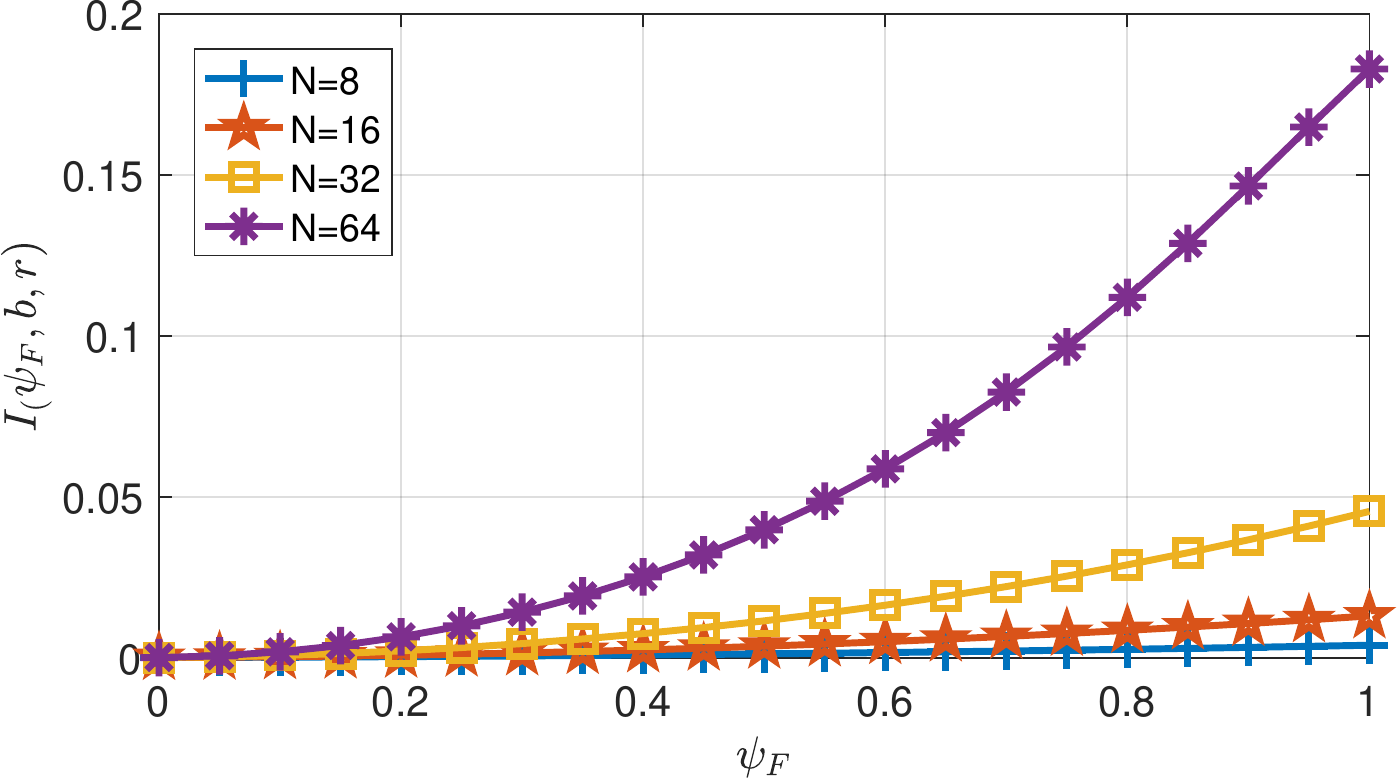}
\caption{Capacity improvement ratio $I\left( {{\psi_F},b, r} \right) $ in a fine beam as a function of beam focus angle $\psi_F$. $r=\frac{\sqrt2}{2}$, $d=\lambda_c/2$, $N_f=2048$, $b=0.0342$ ($B=2.5$ GHz, $f_c=73$ GHz), and $\frac{P}{B\sigma^2}=0$ dB.}
\label{fig:LA_Channel_Capacity_Improvement1_vs_focused_angle_multi_N}
\bigskip
\centering
\includegraphics[width=88 mm]{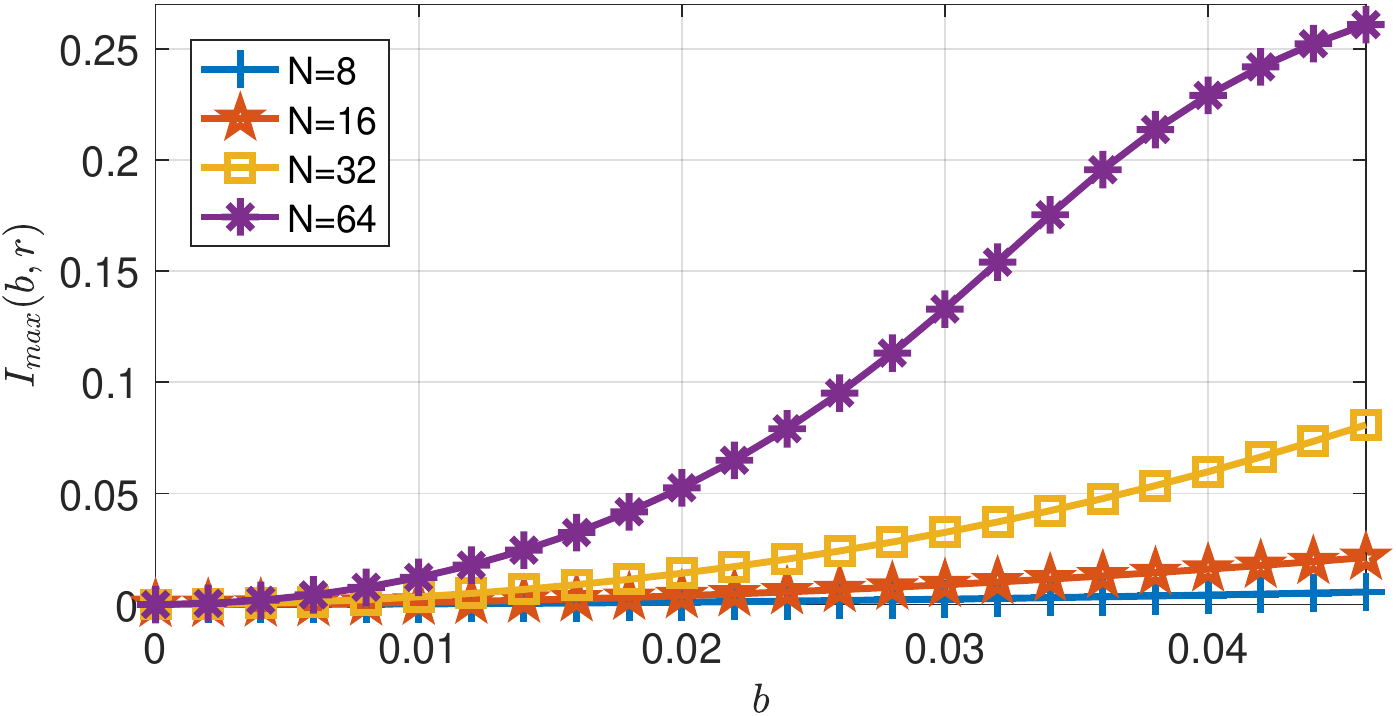}
\caption{Maximum capacity improvement ratio $I_{max}\left( b,r \right) $, as a function of fractional bandwidth $b$. $r=\frac{\sqrt2}{2}$, $d=\lambda_c/2$, $N_f=2048$, and $\frac{P}{B\sigma^2}=0$ dB.}
\label{fig:LA_Channel_Capacity_Improvement2_max_vs_b_multi_N}
\end{figure}

Figure \ref{fig:LA_Channel_Capacity_Improvement1_vs_focused_angle_multi_N} shows that $I\left( {\psi_F},b, \frac{\sqrt2}{2} \right) $ increases as $N$ and $\psi_F$ increase, where $r=\frac{\sqrt2}{2}$ and $\psi \in \mathcal{R}_{3_{\text{dB}}}$. The figure also demonstrates that the effect of beam squint increases as $\psi_F$ increases. Figure \ref{fig:LA_Channel_Capacity_Improvement2_max_vs_b_multi_N} illustrates that ${I_{max}}\left( b, \frac{\sqrt2}{2}  \right)$ increases as $b$ increases, that is, the improvement of the codebook design algorithm will be more significant if the bandwidth is larger. For small $N$, the improvement is negligible.

\subsection{Codebook Size}

\begin{figure}
\centering
\includegraphics[width=88 mm]{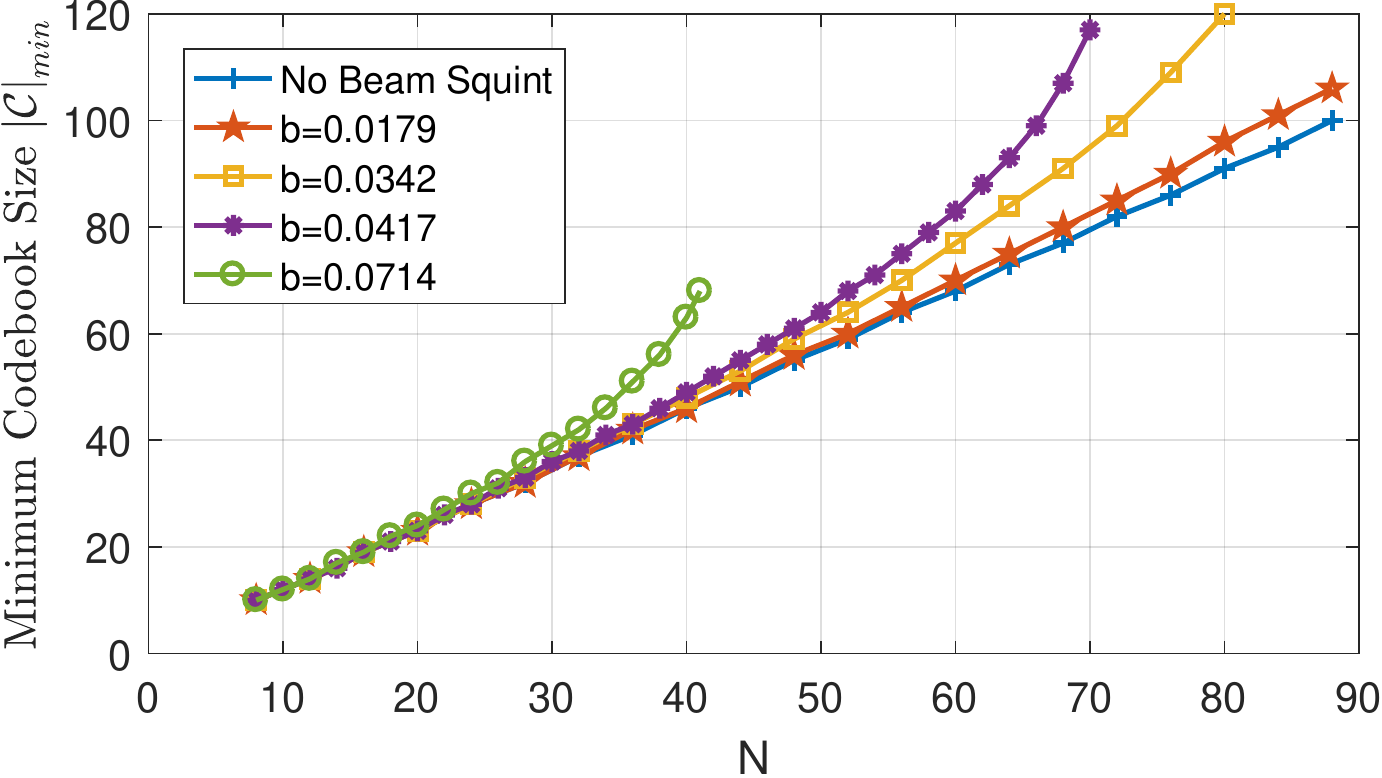}
\caption{Minimum codebook size $\left| \mathcal{C} \right|_{min}$ required to cover $\psi \in \left[ -1, 1 \right]$ as a function of the number of antennas $N$. $C_t$ is set according to (\ref{eq:typical-capacity-threshold}). $d=\lambda_c/2$, $N_f=2048$, and $\frac{P}{B\sigma^2}=0$ dB. $b=0.0179$ corresponds to $B=0.5$ GHz and $f_c=28$ GHz; $b=0.0342$ corresponds to $B=2.5$ GHz and $f_c=73$ GHz; $b=0.0417$ corresponds to $B=2.5$ GHz and $f_c=60$ GHz; $b=0.0714$ corresponds to $B=2$ GHz and $f_c=28$ GHz.}
\label{fig:LA_Codebook_6_Codebook_Size_vs_N_multiple_B_BinaryS}
\end{figure}

Fig.~\ref{fig:LA_Codebook_6_Codebook_Size_vs_N_multiple_B_BinaryS} shows examples of the minimum codebook size $\left| \mathcal{C} \right|_{min}$ as a function of the number of antennas $N$ for $\psi_m=1$, i.e., the maximum covered AoA/AoD of the codebook is $90^{\circ}$. The minimum codebook size increases as $N$ or $b$ increases. Note that the minimum codebook size also varies with $\frac{P}{B\sigma^2}$; however, our simulations suggest that it only has small effect on the codebook size. No codebook exists for $N \ge 42$ with $b=0.0714$.

%

\section{Conclusion}
We analyzed the effect of beam squint as well as its influence on codebook design in analog beamforming with a phased array implemented via phase shifters. Beam squint reduces channel capacity. To compensate for the beam squint, a beamforming codebook is designed with a channel capacity constraint. The codebook size increases as the fractional bandwidth or the number of antennas in the array increases. Furthermore, the fractional bandwidth is upper bounded for a fixed number of antennas.



\bibliographystyle{IEEEtran}
\bibliography{IEEEabrv,mmc}

\begin{thebibliography}{10}
\providecommand{\url}[1]{#1}
\csname url@samestyle\endcsname
\providecommand{\newblock}{\relax}
\providecommand{\bibinfo}[2]{#2}
\providecommand{\BIBentrySTDinterwordspacing}{\spaceskip=0pt\relax}
\providecommand{\BIBentryALTinterwordstretchfactor}{4}
\providecommand{\BIBentryALTinterwordspacing}{\spaceskip=\fontdimen2\font plus
\BIBentryALTinterwordstretchfactor\fontdimen3\font minus
  \fontdimen4\font\relax}
\providecommand{\BIBforeignlanguage}[2]{{%
\expandafter\ifx\csname l@#1\endcsname\relax
\typeout{** WARNING: IEEEtran.bst: No hyphenation pattern has been}%
\typeout{** loaded for the language `#1'. Using the pattern for}%
\typeout{** the default language instead.}%
\else
\language=\csname l@#1\endcsname
\fi
#2}}
\providecommand{\BIBdecl}{\relax}
\BIBdecl

\bibitem{roh2014millimeter}
W.~Roh, J.-Y. Seol, J.~Park, B.~Lee, J.~Lee, Y.~Kim, J.~Cho, K.~Cheun, and
  F.~Aryanfar, ``{Millimeter-Wave Beamforming as an Enabling Technology for
  {5G} Cellular Communications: Theoretical Feasibility and Prototype
  Results},'' \emph{IEEE Commun. Mag.}, vol.~52, no.~2, pp. 106--113, 2014.

\bibitem{rappaporttheodores2002}
T.~S. Rappaport, \emph{{Wireless Communications: Principles and Practice}},
  2nd~ed.\hskip 1em plus 0.5em minus 0.4em\relax Upper Saddle River, New
  Jersey: Prentice Hall, 2002.

\bibitem{heath2015overview}
R.~W. Heath, Jr., N.~Gonzalez-Prelcic, S.~Rangan, W.~Roh, and A.~Sayeed, ``{An
  Overview of Signal Processing Techniques for Millimeter Wave {MIMO}
  Systems},'' \emph{IEEE J. Sel. Topics Signal Process.}, vol.~10, pp.
  436--453, 2016.

\bibitem{song2015codebook}
J.~Song, J.~Choi, and D.~J. Love, ``{Codebook Design for Hybrid Beamforming in
  Millimeter Wave Systems},'' in \emph{Proc. 2015 IEEE Int. Conf. on Commun.
  (ICC)}.\hskip 1em plus 0.5em minus 0.4em\relax IEEE, 2015, pp. 1298--1303.

\bibitem{mailloux2005phased}
R.~J. Mailloux, \emph{{Phased Array Antenna Handbook}}, 2nd~ed.\hskip 1em plus
  0.5em minus 0.4em\relax Norwood, MA: Artech House, 2005.

\bibitem{SeyedGarakkoui2011BeamSquinting}
S.~Garakoui, E.~Klumperink, B.~Nauta, and F.~E. van Vliet, ``{Phased-Array
  Antenna Beam Squinting Related to Frequency Dependence of Delay Circuits},''
  in \emph{Proc. 41th European Microwave Conf.}, Manchester, UK, Oct. 2011.

\bibitem{cai2016effect}
M.~Cai, K.~Gao, D.~Nie, B.~Hochwald, J.~N. Laneman, H.~Huang, and K.~Liu,
  ``{Effect of Wideband Beam Squint on Codebook Design in Phased-Array Wireless
  Systems},'' in \emph{2016 IEEE Global Commun. Conf. (GLOBECOM)}.\hskip 1em
  plus 0.5em minus 0.4em\relax Washington, DC: IEEE, Dec. 2016.

\bibitem{longbrake2012true}
M.~Longbrake, ``{True Time-Delay Beamsteering for Radar},'' in \emph{2012 IEEE
  National Aerospace and Electronics Conference (NAECON)}.\hskip 1em plus 0.5em
  minus 0.4em\relax IEEE, 2012, pp. 246--249.

\bibitem{cover2012elements}
T.~M. Cover and J.~A. Thomas, \emph{{Elements of Information Theory}},
  2nd~ed.\hskip 1em plus 0.5em minus 0.4em\relax Hoboken, New jersey: John
  Wiley \& Sons, 2006.

\bibitem{thomas2001introduction}
T.~H. Cormen, C.~E. Leiserson, R.~L. Rivest, and C.~Stein, \emph{{Introduction
  to Algorithms}}, 3rd~ed.\hskip 1em plus 0.5em minus 0.4em\relax Cambridge,
  MA: MIT Press, 2009.

\end{thebibliography}

\end{document}